# Toward a Comprehensive Model of Snow Crystal Growth:
# 7. Ice Attachment Kinetics near -2 C

Kenneth G. Libbrecht

Department of Physics
California Institute of Technology
Pasadena, California 91125
kgl@caltech.edu

**Abstract. I examine a variety snow crystal growth experiments performed at temperatures near -2 C, as a function of supersaturation, background gas pressure, and crystal morphology. Although the different experimental data were obtained using quite diverse experimental techniques, the resulting measurements can all be reasonably understood using a single comprehensive physical model for the basal and prism attachment kinetics, together with particle diffusion of water vapor through the surrounding medium and other well-understood physical processes. As with the previous paper in this series, comparing and reconciling different data sets at a single temperature yields significant insights into the underlying physical processes that govern snow crystal growth dynamics.**

## 1. Introduction

As suggested by its title, this series of papers represents my ongoing efforts to develop a comprehensive model of the physical dynamics of snow crystal growth. The research is very much a work in progress, as the formation of ice crystal structures from the solidification of water vapor is a remarkably rich and subtle phenomenon, involving the complex interplay of many physical processes. Developing a suitable comprehensive model has required, and continues to require, constant adaptations as additional information is provided by new theoretical ideas and new experimental observations taken over a range of environmental conditions. Some past reviews of this subject can be found in [1954Nak, 1987Kob, 2005Lib, 2017Lib, 2019Lib].

In a recent paper [2019Lib1], I described a novel semi-empirical physical model for the snow-crystal attachment kinetics, describing the attachment coefficients $\alpha_{basal}$ and $\alpha_{prism}$ as a function of supersaturation and temperature from approximately -1 C to -30 C. The model appears to reproduce most of the salient features of the snow-crystal morphology diagram over this substantial temperature range, a phenomenon that has been largely unexplained for nearly 70 years. Moreover, the model is a fully quantitative, making many testable predictions of ice growth rates over a broad range of conditions, using plausible assumptions regarding the molecular dynamics of terrace nucleation, surface diffusion, and other physical effects. While the model is certainly speculative to some degree, so far it seems to be holding up to critical examination using precision ice-growth measurements, and the present paper provides additional confirmation in this regard.

In the previous paper in this series [2019Lib2], I found much benefit in examining ice growth at -5 C using an assorted collection of new and previously published experimental data. This specific temperature was especially enlightening, as both plate-like and columnar morphologies can be readily observed under different conditions. Knight [2012Kni] even demonstrated that these two extreme morphologies can co-develop simultaneously under certain conditions, indicating a subtle sensitivity to initial conditions. As described in [2019Lib2], the model in [2019Lib1] nicely explains this peculiar set of seemingly disparate growth behaviors, which is otherwise quite difficult to comprehend.

I continue this line of reasoning here by examining ice-growth data at -2 C as a function of supersaturation and other conditions. In a similar fashion, I hope to convince the reader that the different experimental results are converging on a set of well-defined growth behaviors under different environmental and initial conditions. Moreover, the aforementioned model does quite a reasonable job of explaining the entire data collection.

## 2. A Physical Model of the Ice Attachment Kinetics

Because much of the needed background information has already been provided in prior publications [2019Lib, 2019Lib1, 2019Lib2], I proceed directly into a description of my proposed model of the basal and prism attachment kinetics at -2 C. Figure 1 illustrates the model attachment coefficients $\alpha_{basal}$ and $\alpha_{prism}$ as a function of $\sigma_{surf}$ over typical growth conditions near -2 C, while Figure 2 shows the same model converted to growth velocities $v_{basal}$ and $v_{prism}$. The two figures are related via the kinetic growth equation

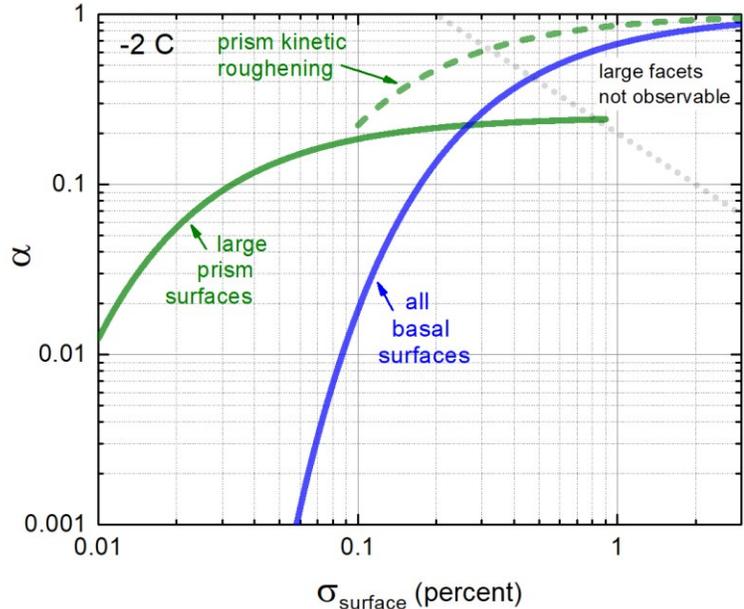

**Figure 1:** The proposed model of the ice/vapor attachment coefficients on basal and prism surfaces when the growth temperature is near -2 C, adapted from [2019Lib1]. The physical significances of the different curves are discussed in detail in the text. Similar curves at -5 C are described in [2019Lib2].

$$v_n = \alpha v_{kin} \sigma_{surf} \quad (1)$$

with $v_{kin} \approx 635 \ \mu m/sec$ at -2 C [2019Lib].

Note that the full temperature-dependent model of the attachment kinetics has been described in [2019Lib1], and Figure 1 shows a detailed look at this model for a fixed temperature of -2 C. A single-valued function describes $\alpha_{basal}(\sigma_{surf})$ over the entire range of $\sigma_{surf}$ shown, but the growth of faceted prism surfaces requires two separate "branches" of $\alpha_{prism}(\sigma_{surf})$, and these are described in detail below. I assume that the attachment kinetics on non-faceted ice surfaces are described by $\alpha \approx 1$ at all $\sigma_{surf}$.

### Basal Attachment Kinetics

Consider first the blue curve labeled "all basal surfaces" in Figure 1, which represents the growth of faceted basal surfaces. In this case the attachment kinetics are well described by the nucleation and growth of terrace steps, which has a well-established theory discussed in most crystal-growth textbooks [1996Sai, 1999Pim, 2002Mut]. Assuming a poly-nucleation model in classical nucleation theory from the vapor phase, I write the attachment coefficient as

$$\alpha_{basal}(\sigma_{surf}) = A_{basal} e^{-\sigma_{0,basal}/\sigma_{surf}} \quad (2)$$

where $\sigma_{0,basal}$ is a parameter that derives from the terrace step energy on the faceted surface and $A_{basal}$ depends on the admolecule surface diffusion and other parameters.

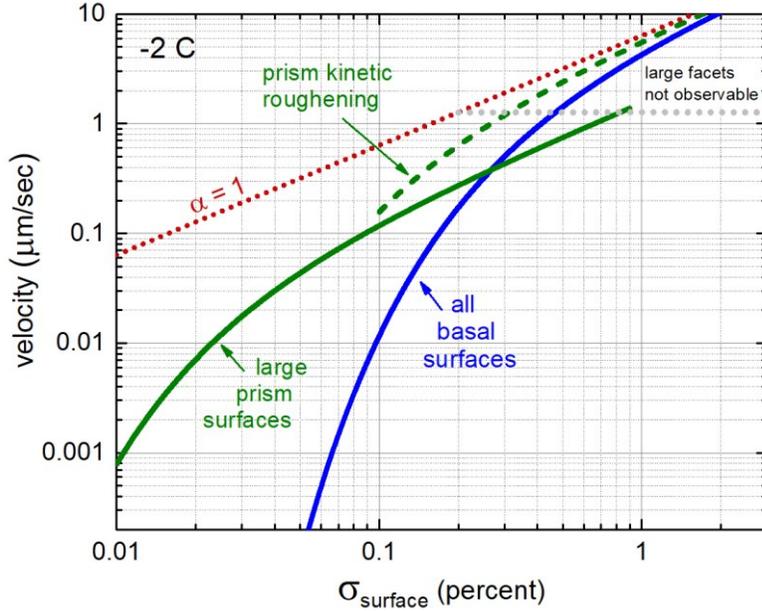

**Figure 2:** (left) The model in Figure 1 converted to growth velocities using Equation (1).

In classical nucleation theory, $A_{basal}$ may exhibit a weak dependence on $\sigma_{surf}$, but this dependence is negligibly small in the current discussion, given the substantial uncertainties in experimental data. Thus I assume that both $\sigma_{0,basal}$ and $A_{basal}$ are constant at -2 C, independent of $\sigma_{surf}$. Measurements indicate (see below) that $\sigma_{0,basal} \approx 0.004 = 0.4\%$ and $A_{basal} \approx 1$ so these values were used to define the basal curve in Figure 1.

Although the phenomenon of Structure-Dependent Attachment Kinetics (SDAK) played an important role in the basal growth at -5 C, it appears to be essentially absent at -2 C. In that regard, the model behavior at -2 C is relatively simpler compared to that at -5 C. The experimental data suggest that a simple nucleation model applies to both broad and narrow basal facets, unlike at -5 C. An experimental investigation into the transition in growth behaviors between -5 C and -2 C could yield interesting physical insights into the basal attachment kinetics and the SDAK phenomenon, but this is beyond the scope of the present paper.

Note that it is generally not possible to observe the growth of broad faceted structures at especially high growth rates, as is indicated in Figures 1 and 2. In this high-growth-velocity regime, thermal and particle diffusion become dominant growth-limiting factors, resulting in the formation of dendritic structures that do not possess broad faceted features. Nevertheless, there must always be a faceted basal surface on such structures, defined by the top basal terrace on an otherwise curved dendrite tip. In the model defined by Figure 1, the growth of any basal surface is governed by the nucleation of additional terraces on the faceted plane, regardless of the lateral dimensions of the top terrace. This statement is consistent with the available experimental evidence, as measurements generally suggest (see below) $\alpha_{basal}(\sigma_{surf}) \to 1$ in the limit $\sigma_{surf} \gg \sigma_{0,basal}$ [2013Lib], supporting the parameter choice of $A_{basal} = 1$.

### Prism Attachment Kinetics

Turning our attention to the prism facet, the "large prism surfaces" curve in Figure 1 is again defined by a nucleation-limited model, this time using $\sigma_{0,prism} = 0.03\%$ and $A_{prism} = 0.25$ at -2 C. As the label suggests, this curve applies when the faceted prism surfaces have large lateral dimensions, typically meaning greater than roughly 1-2 $\mu$m. This usually also means that the prism growth velocities are relatively slow, less than about 1 $\mu$m/sec.

Taking a broader perspective for a moment, terrace nucleation appears to dominate the ice/vapor attachment kinetics on large basal and prism facets over quite a broad range of temperatures [2013Lib]. Basal growth at all temperatures is consistent with $A_{basal} \approx 1$, while $A_{prism} \approx 1$ applies only to temperatures below -10 C. Above this temperature, the growth of large prism facets is better described with $A_{prism} < 1$ [2013Lib].

The decrease in $A_{prism}$ at higher temperatures appears to be related to the development of a substantial quasi-liquid layer (QLL) on the prism surface above -10 C. The physical dynamics underlying the observed change in $A_{prism}$ with temperature is necessarily quite complex, in part because terrace nucleation must occur at the QLL/ice interface when the QLL is thick. The theory of crystalline attachment kinetics at liquid/solid interfaces is overall not as well developed as at vapor/solid interfaces, so clearly QLL/solid interfaces will be even more difficult to comprehend. For the present, therefore, I take $A_{prism} < 1$ on large prism facets at higher temperatures to be a largely unexplained empirical fact.

For smaller facet dimensions and/or faster growth velocities, the prism behavior transitions to the "prism kinetic roughening" curve in Figure 1, defined by $A_{prism} = 1$ and $\sigma_{0,prism} = 0.15\%$. Although drawn as a well-defined curve in Figure 1, this should be considered as a rough approximation for a poorly understood growth regime. The physical nature of this branch likely has little connection to terrace nucleation theory, in spite of the chosen functional form. I speculate that this behavior may be related to the phenomenon of kinetic roughening [1996Sai, 1999Pim, 2002Mut], but that is little more than a guess at this point.

The inclusion of this fast-growth branch in the prism kinetics model was motivated by the available data (below) that suggests $\alpha_{prism}(\sigma_{surf}) \to 1$ on fast-growing dendrite tips at high supersaturations. This behavior appears to be qualitatively different from the growth of broad faceted prism surfaces, so it is represented by a separate branch in Figure 1.

The transition between these two behaviors has received little study, so the two prism branches overlap in the region $0.1 < \sigma_{surf} < 1$ percent. This is meant to indicate that the transition from one curve to the other will happen somewhere in this region, and it may be smooth or abrupt; the exact behavior is not yet constrained by either theory or experiment. The growth behaviors at low $\sigma_{surf}$ and high $\sigma_{surf}$ appear to follow their respective branches in Figure 1, but the intermediate region is not well defined.

Some additional caveats and physical considerations are presented in [2019Lib2] for the -5 C case, and these largely apply at -2 C as well. In particular, I ignore any dependence of the attachment kinetics on background gas pressure (for air and other similarly inert gases), as there is little evidence for gas-related effects in the data. Of course, water-vapor diffusion through the background gas is often a substantial effect that limits growth rates, but this is a separate issue from the attachment kinetics.

# 3. Comparisons with Ice Growth Data

Paralleling the discussion presented in [2019Lib2], we now compare the quantitative model in Figure 1 with relevant experimental data. Our overarching goal is to examine all available measurements acquired using different techniques under a variety of environmental conditions, in order to better understand the full range of physical processes that affect ice growth dynamics, including attachment kinetics, particle diffusion, latent heating, surface diffusion, and surface-energy effects. But our focus here is mainly on the attachment kinetics of basal and prism surfaces.

## Growth on Substrates at Low Background Pressures

The most useful measurements for examining the attachment kinetics are those taken at low background gas pressure, as diffusion effects are generally smaller at lower pressures, making it easier to accurately determine $\sigma_{surf}$, and thus allowing precise measurements of $\alpha_{basal}$ and $\alpha_{prism}$ from observed growth velocities.

Figure 3 shows measurements taken by Libbrecht and Rickerby using the experimental apparatus described in [2013Lib]. I believe that this experiment provided one of the most precise measurements of the ice/vapor attachment kinetics, substantially superseding previous efforts, as it covered a wide range in $T$ and $\sigma_{surf}$ while carefully managing a variety of systematic experimental effects. While fit parameters as a function of $T$ were reported in [2013Lib], Figure 3 shows less-processed data at -2 C from that experiment.

Figure 4 shows additional data obtained using the newer experimental apparatus described in [2019Lib3]. These two experiments were similar in character, as each measured small crystals (typically 3-30 μm in size) at low background pressures (30-60 mbar) growing on sapphire substrates. Both used plane-parallel growth chambers with ice reservoirs in close proximity to the growing test crystals, thus allowing accurate modeling of residual diffusion effects to extract $\sigma_{surf}$ from known experimental parameters [2019Lib, 2019Lib3].

There were also several substantial differences between these two experiments. In [2013Lib], for example, substrate interactions were avoided by measuring the growth of faceted surfaces aligned parallel to the substrate using white-light interferometry. Because these surfaces did not intersect the substrate, there was no possibility of enhanced terrace nucleation from unwanted substrate interactions [2019Lib]. In contrast, substrate interactions in [2019Lib3] were reduced by use of a hydrophobic coating on the sapphire surface, which seemed to provide a suitably large ice/substrate contact angle that reduced spurious terrace nucleation. Chemical-vapor contamination effects were likely smaller in [2019Lib3], and larger numbers of crystals were observed in this experiment to examine reproducibility and other systematic effects. The newer experiment also observed smaller crystals, with careful modeling of the diffusion field to extract $\sigma_{surf}$ from the data [2019Lib3].

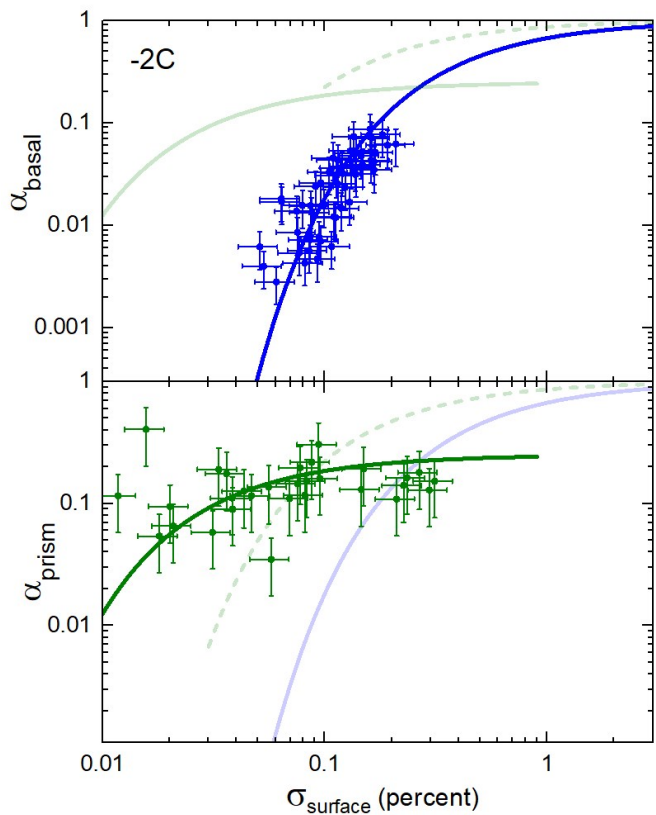

**Figure 3:** Measurements of $\alpha_{basal}$ (top, blue data points) and $\alpha_{prism}$ (bottom, green data points) as a function of $\sigma_{surf}$, at a growth temperature of $T \approx -2\ C$. The data are from the experiment described in [2013Lib], taken at background air pressures near 0.03 bar. The lines are reproduced from Figure 1.

Comparing Figures 3 and 4, it is immediately clear that the two experiments yielded remarkably similar results. Although they were performed at different times, using entirely different hardware and somewhat different experimental strategies, the overall data trends for both $\alpha_{basal}(\sigma_{surf})$ and $\alpha_{prism}(\sigma_{surf})$ show excellent agreement. In both experiments, much attention was given to understanding and modeling diffusion effects and other systematic errors that could affect the measurements. The model curves in Figure 1 for large faceted surfaces were largely derived from these low-pressure data.

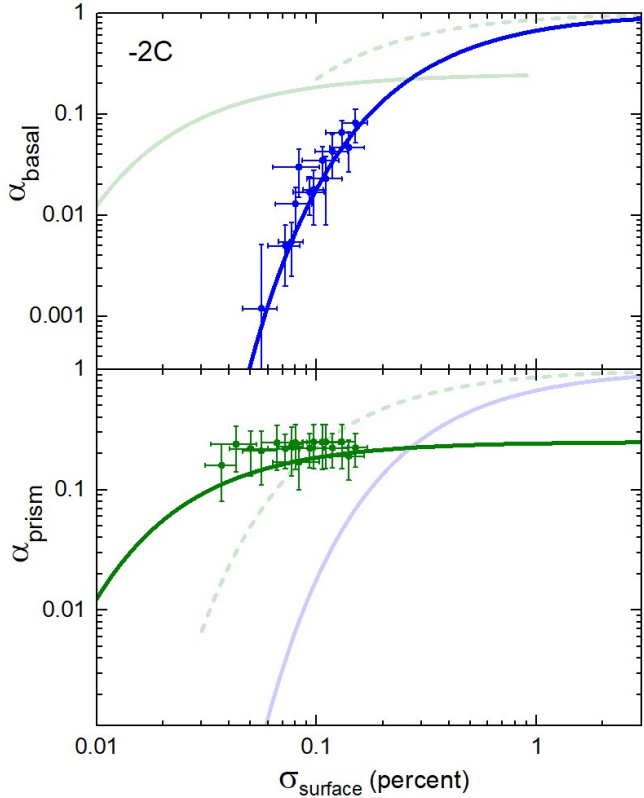

**Figure 4: Additional measurements of $\alpha_{basal}$ (top, blue) and $\alpha_{prism}$ (bottom, green) as a function of $\sigma_{surf}$ at a growth temperature of $T \approx -2\ C$ and a background air pressure of 0.06 bar. These data were obtained using the newer substrate-growth experiment described in [2019Lib3], and the lines are again reproduced from Figure 1. Note that the new data, from a completely redesigned ice-growth experiment, are in good agreement with the older data shown in Figure 3.**

The basal data are clearly well fit using a nucleation model with $A_{basal} \approx 1$, and [2013Lib] found this to be the case for all temperatures between -2 C and -30 C. The basal step energy, and therefore $\sigma_{0,basal}$, changes substantially with temperature over this range, but $A_{basal} \approx 1$ provided a good fit over the entire data set. This was not true for the growth of prism facets, where $A_{prism} \approx 1$ provided a good fit at temperatures below -10 C, but lower values of $A_{prism}$ were needed to fit the data at higher temperatures.

If we try to fit either of the data sets in Figures 3 and 4 while constraining the fit to have $A_{prism} = 1$, we cannot obtain satisfactory agreement with the observations from either experiment. The light-green dashed curves in Figures 3 and 4 are about the best one can do with a constrained model, and in both data sets these curves yield substantially stronger changes in $\alpha_{prism}(\sigma_{surf})$ with $\sigma_{surf}$ than are shown in the data. Taken together, I believe that these two experiments provide quite convincing evidence that a terrace nucleation model with $A_{prism} \approx 1$ cannot describe the growth of large prism facets at temperatures near -2 C.

Early experimental measurements of ice growth rates in near-vacuum were reported by Lamb and Scott [1972Lam], who examined a substantial range of temperatures. Although -2 C is near the end of their measurement range, they observed growth rates of about 1 $\mu$m/sec for both basal and prism facets at $\sigma_{surf} \approx 0.25\%$. These velocities are roughly in agreement with the model in Figure 2, although the measurement uncertainties were fairly

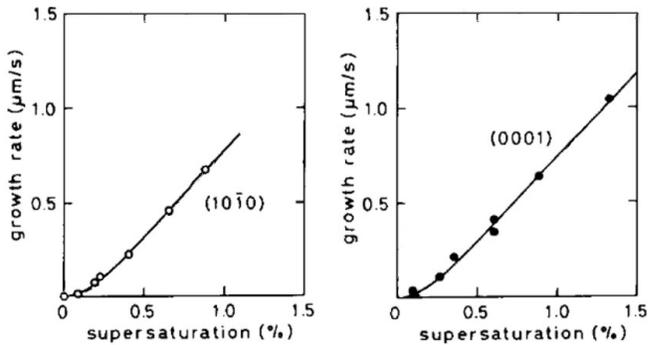

**Figure 5: Early measurements of ice growth rates at -1.9 C in a low background pressure, for prism (left) and basal (right) surfaces [1989Sei].**

large, and the model curves are quite closely spaced at such high velocities.

Figure 5 shows additional near-vacuum measurements from Sei and Gonda [1989Sei] at a temperature of -1.9 C, showing (for example) growth rates of about 0.25 $\mu$m/sec for both basal and prism facets at $\sigma_{surf} \approx 0.5\%$. The data are somewhat off from the model but show reasonable agreement if we reduce the assumed supersaturation to $\sigma_{surf} \approx 0.25\%$. Residual diffusion effects could be responsible for this reduction, as this is a difficult systematic error to eliminate entirely. I believe that the experiments described in Figures 3 and 4, learning from previous efforts, used improved chamber designs to realize more accurate estimates for $\sigma_{surf}$. As I discussed in [2004Lib], systematic errors in estimating $\sigma_{surf}$ were a significant problem in many of the earlier ice-growth experiments.

## Free-Fall Growth in Air

Figure 6 presents another relevant data set taken at -2 C, showing measurements of crystals that grew in free-fall through supersaturated air at a pressure of 1 bar [2008Lib1]. In essence, a vessel of heated water at the bottom of a one-meter-tall growth chamber yielded supersaturated air as convection mixed the evaporated water vapor with the air in the chamber. The resulting supersaturation was not well determined, and Figure 6 shows $\sigma_\infty$ values measured using differential hygrometry at the center of the chamber. Subsequent analysis suggested that these quoted $\sigma_\infty$ values from [2008Lib1] were too high by perhaps a factor of two, and generally we have found that such low supersaturation levels are quite difficult to measure directly with reasonable accuracy.

In addition to a substantial uncertainty in $\sigma_\infty$, diffusion corrections make it exceedingly difficult to determine $\sigma_{surf}$ accurately in these free-fall measurements. This is best understood from the analytical model for spherical growth [2019Lib], which indicates

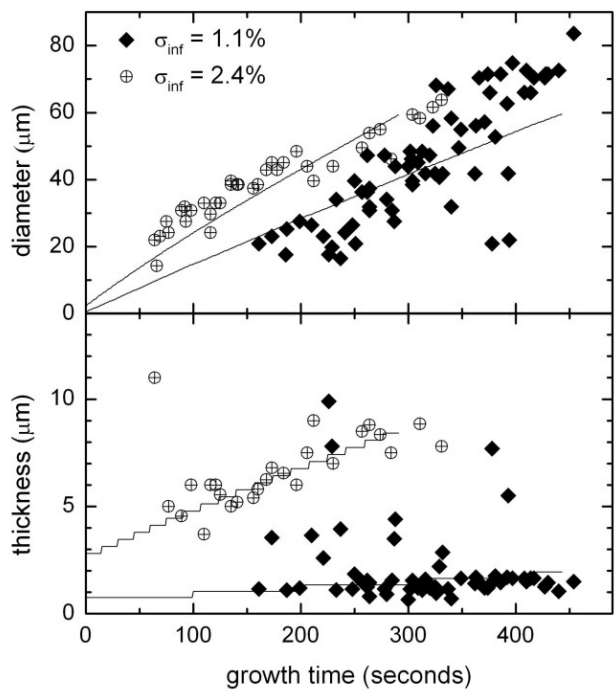

**Figure 6: Free-fall growth data in normal air at -2 C from [2008Lib1], showing crystal sizes as a function of fall times for two different supersaturations. Each pair of points (diameter and thickness) represent one observed crystal, and lines show growth models described in [2008Lib1]. The $\sigma_\infty$ values give the estimated supersaturations far from the growing crystals, which is generally substantially higher than $\sigma_{surf}$ because of diffusion effects. From these data I extract two data points having $(v_{basal}, v_{prism}) \approx (12, 100)$ nm/sec and $\approx (1.2, 67)$ nm/sec.**

$$\delta\sigma = \sigma_\infty - \sigma_{surf}$$
$$\approx \frac{R}{X_0} \frac{v}{v_{kin}} \qquad (3)$$

Using typical crystal sizes and growth velocities from Figure 6 quickly reveals that $\delta\sigma$ is not a small correction relative to $\sigma_\infty$, meaning that substantial experimental uncertainties in both $\sigma_\infty$ and $\delta\sigma$ yield large uncertainties in $\sigma_{surf}$. This fact was not sufficiently appreciated in [2008Lib1], and I now believe that the analysis for $\alpha_{basal}$ and $\alpha_{prism}$ in that paper was largely incorrect.

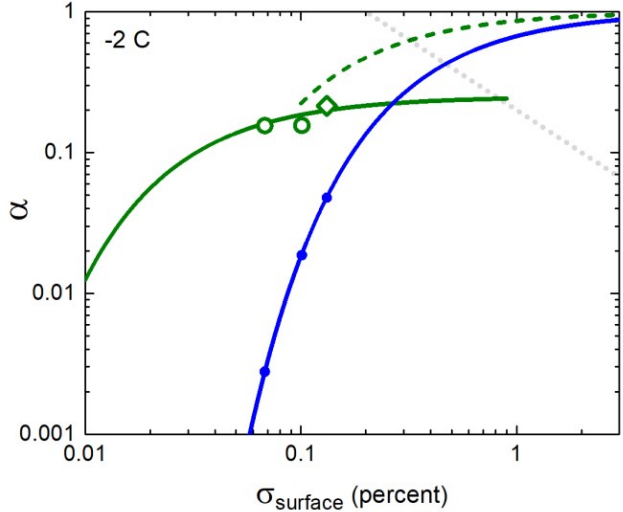

**Figure 7: A comparison of the free-fall data with the model in Figure 1 using data from [2008Lib1] (circles) and [1987Kob] (diamond). Here the measured basal growth velocity was assumed to agree with the model (solid blue points), thus determining $\sigma_{surf}$ on the basal surface, and this value was assumed to be equal on the basal and prism surfaces. The values of $\alpha_{prism}$ (open green points) were then calculated from $\sigma_{surf}$ and the measured prism growth velocities.**

To obtain some useful information from these free-fall data, therefore, I use a "witness surface" analysis by assuming that the basal growth rates are determined exactly by the $\alpha_{basal}$ model described in Figures 1 and 2. Using the measured growth velocities from Figure 6 together with the basal velocity curve in Figure 2 then yields a direct estimate of $\sigma_{surf}$ at the basal surfaces. A numerical diffusion model further suggests that this same value of $\sigma_{surf}$ applies to the nearby prism surfaces, as the test crystals are quite small. Assuming a near-constant $\sigma_{surf}$ on the crystal, each measured prism growth velocity then yields a measurement of $\alpha_{prism}$ from Equation 1.

The two sets of data in Figure 6 thus yielded the two data points for $\alpha_{prism}(\sigma_{surf})$ shown in Figure 7. Although the analysis is model-dependent and somewhat crude, the points nevertheless show reasonably good agreement with the model. In particular, the model predicts that the crystal aspect ratio (diameter/thickness) should change rapidly with supersaturation, and this is observed in the data. The low inferred values of $\sigma_{surf}$ support the above analysis indicating that the diffusion corrections are large and difficult to model in an absolute sense from the data. The "witness surface" approach is therefore about the best one can to with these data.

Note that Figure 7 also supports our assumption that the attachment kinetics does not depend strongly on the background air pressure, and additional evidence of this hypothesis at -5 C was presented in [2019Lib2]. At least in this high-temperature regime, the evidence suggests that the ice attachment kinetics in air is essentially independent of pressure between 0.05 bar and 1 bar.

Figure 8 shows additional free-fall growth measurements in a cloud chamber from Yamashita [1987Kob], where the supersaturation was held near $\sigma_\infty \approx \sigma_{water} \approx 2\%$ by a cloud of freely-floating water droplets in the chamber. Taking $(v_{basal}, v_{prism}) \approx (40, 180)$ nm/sec at -2 C from this graph, and applying the same witness-surface analysis, we again find quite good agreement with our model, as shown in Figure 7.

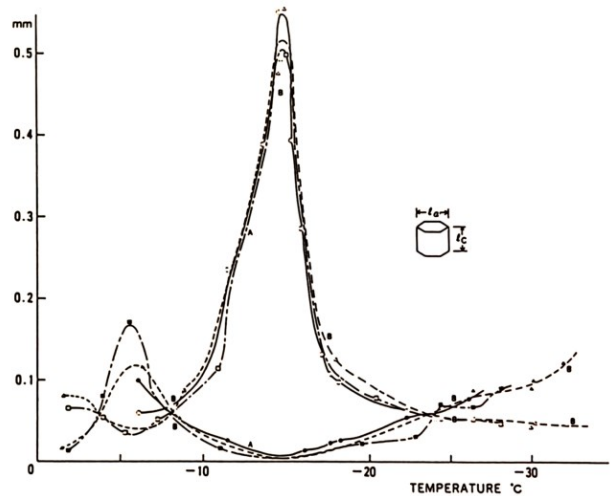

**Figure 8: Measurements of the diameters and thicknesses of snow crystals after growing in air with $\sigma_\infty \approx \sigma_{water} \approx 2\%$ for 200 seconds in a free-fall cloud chamber [1987Kob].**

## DENDRITIC STRUCTURES IN AIR

The growth of dendritic structures on c-axis ice needles ("electric" ice needles in [2019Lib]) provides another interesting experimental system that can be employed to test our model of the attachment kinetics. Briefly, high electric fields are used to stimulate the growth of c-axis needle crystals in the first of two diffusion chambers, and the needles are then transported to a second diffusion chamber to observe their subsequent growth in normal air with no applied electric fields [2014Lib1, 2019Lib].

This apparatus allows reproducible observations at quite high supersaturation levels in air, although the growth of the resulting large structures is strongly affected by diffusion effects.

Figure 9 shows several representative observations of the growth of prismatic, platelike, and dendritic structures on the ends of thin ice needles in air at -2 C. These and similar images illustrating growth at 20 different temperatures from -0.5 C to -21 C, each at these same supersaturation levels, can be found in [2019Lib].

It is instructive to first examine the tip growth of the high-$\sigma_\infty$ dendritic structures in Figure 9, analyzing multiple images of growing crystals to extract the component growth velocities $v_{basal}$ and $v_{prism}$ for each dendrite tip. Once again, I use a witness-surface analysis by assuming the basal model in Figure 2 is correct and then using each measured $v_{basal}$ value to estimate $\sigma_{surf}$ at the corresponding dendrite tip. Assuming that this same $\sigma_{surf}$ applies to the nearby prism surfaces on the same tip, the measured $v_{prism}$ then gives an estimate of $\alpha_{prism}(\sigma_{surf})$, as described with the analysis of the free-fall data above. Again, the analysis is clearly model-dependent, but it does provide some useful information at high growth velocities that is otherwise difficult to obtain.

Results from this witness-surface analysis of dendrite tip growth are shown in the fast-growing data points in Figure 10. Three of these points come from the highest-$\sigma_\infty$ crystals in Figure 9, while the other two points are from similar data not shown. The crystal morphologies seen in Figure 9 immediately indicate that $\alpha_{prism} > \alpha_{basal}$ on the uppermost basal and prism terraces on the dendrite tips, and Figure 10 further suggests that $\alpha_{prism} \to 1$ at very high $\sigma_{surf}$. As described in the model section above, I take this behavior to be indicative of some kind of kinetic-roughening behavior when the prism growth rate is especially high, and/or the prism facet width is especially small. The physical nature of this behavior is not known at present, but these data clearly indicate that the dendrite tip growth is not well described by the "large prism surfaces" curve in Figure 1.

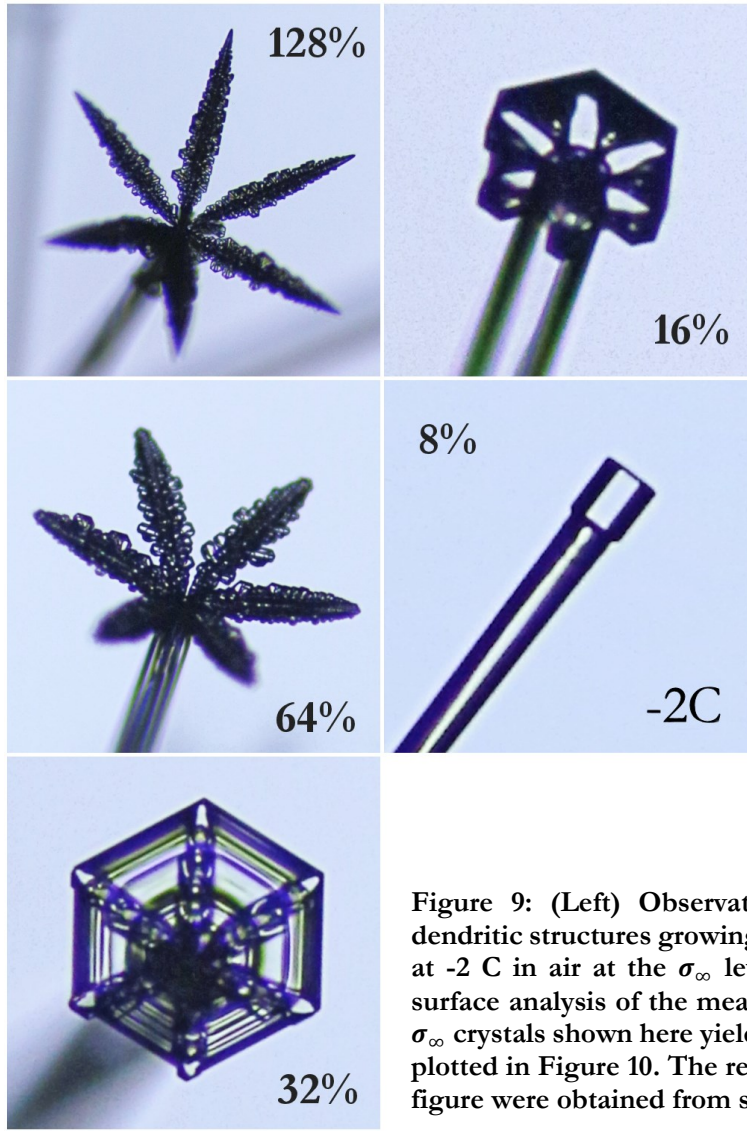

**Figure 9:** (Left) Observations of the growth of columnar and dendritic structures growing on the ends of c-axis ice needle crystals at -2 C in air at the $\sigma_\infty$ levels shown [2019Lib]. Using a witness-surface analysis of the measured tip velocities of the three highest-$\sigma_\infty$ crystals shown here yielded three of the high-velocity data points plotted in Figure 10. The remaining two high-velocity points in that figure were obtained from similar imaging data (not shown here).

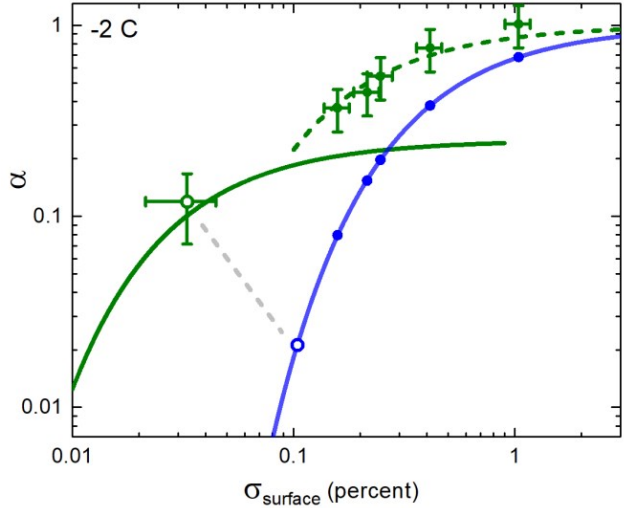

**Figure 10:** The five fast-growing points in this plot show a witness-surface analysis applied to the growth of dendrite tips, as described in the text. The low-velocity point (indicated by a grey dotted line) refer to the growth of a simple needle crystal, which is also described in the text.

### SLENDER ICE NEEDLES IN AIR

The growth of c-axis needles in air at low supersaturation presents another interesting test case, which can be illustrated by the data shown in Figures 11 and 12. For this crystal, the simple witness-surface analysis described previously does not work, in part because the growing surfaces are so large, and in part because the negative needle taper (with the tip thicker than the base) gives a vicinal surface with $\alpha_{vicinal} \approx 1$ over most of the sides of the needle [2019Lib].

To proceed, therefore, I modeled this needle using the cylindrically symmetric cellular automata model described in [2013Lib1, 2019Lib]. The model assumed an $\alpha_{basal}(\sigma_{surf})$ equal to that shown in Figure 1 together with $\alpha_{prism} \approx \alpha_{vicinal} = 1$. Details of this model analysis will be provided in a later paper, but Figure 13 shows the essential result.

The main take-away from this result is that there is a large difference between $\sigma_{surf}$ on the basal needle-tip surface and $\sigma_{surf}$ along the vicinal sides of the needle. This happens because the crystal dimensions are quite large (compared to the characteristic diffusion length $X_0 \approx 0.15\,\mu m$ in air [2019Lib]), together with the fact that $\alpha_{vicinal} \approx 1$ along most of the sides of the needle.

Note that the top prism terrace only exists on a narrow ring quite near the tip of the needle, because what appears to be a flat prism facet surface in Figure 11 likely has a slight negative taper, not measurable in the optical images. For this specific needle example, we see that $\sigma_{surf}$ at the top basal terrace is roughly 3x larger than $\sigma_{surf}$ at the top prism terrace, which is a substantial difference. The model was adjusted to give roughly the observed $v_{basal}$, and the 3x lower $\sigma_{surf}$ was used to estimate $\alpha_{prism}$, giving the low-$\sigma_{surf}$ data point shown in Figure 10. Although quite crude, the numerical analysis suggests that this needle example is at least not inconsistent with the model in Figure 1.

The potential for model-dependent systematic errors in this analysis is clearly large, so the low-$\sigma_{surf}$ data point shown in Figure 10 cannot be considered a solid result, even with the large error bars shown. Nevertheless, I present this example because it provides several insights into the overall problem of understanding snow crystal growth. First, the rough agreement with Figure 1 provides a consistency check; there are no showstoppers here that would cause us to reject the kinetics model we adopted at the beginning of this paper. Second, the analysis reveals some of the challenges inherent in modeling the growth of large structures in air, especially when vicinal surfaces with $\alpha_{vicinal} \approx 1$ are present. Numerical growth models struggle to reproduce faceted crystal growth in general [2019Lib], and this example illustrates that mixtures of vicinal and faceted surfaces may be especially useful for developing and testing these computational techniques. Third, this example suggests new experiments comparing the growth of fully faceted and vicinal surfaces. The electric-needle technique seems well suited to this research direction, as it can provide both positive-taper, negative-taper, and faceted needle crystals. I suspect that additional measurements with this system may lead to interesting insights into the general problem of structure formation in faceted crystal growth.

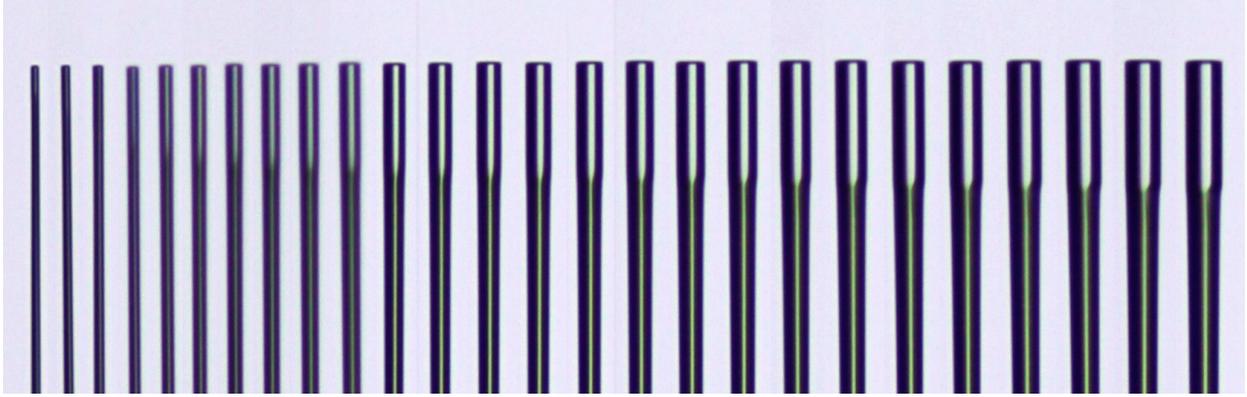

**Figure 11: A series of images showing the growth of a single c-axis ice needle in air at -2 C with $\sigma_\infty \approx 3.7\%$. A cluster of slowly growing frost crystals at the needle base (not shown in these cropped images) provided a stable "base reference" that allowed the axial growth rate to be measured with good accuracy. Radial growth was measured simply from the columnar width at the needle tip. Figure 12 shows the measured needle height and radius as a function of time extracted from these images.**

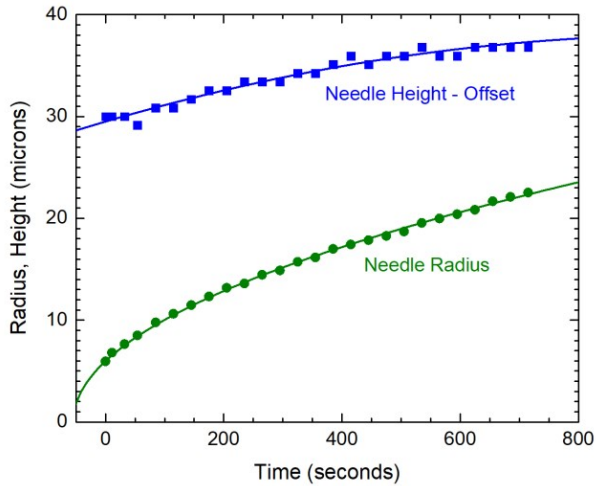

**Figure 12: Measurements of the needle height $H(t)$ and tip radius $R(t)$ extracted from the images in Figure 11. Note that an offset constant has been subtracted from $H(t)$. Lines were drawn to guide the eye, giving $(v_{basal}, v_{prism}) \approx (14, 25)$ nm/sec at $t = 200$ seconds.**

## 4. Conclusions

As with the previous paper in this series [2019Lib2], it appears that the available experimental data at -2 C provide good support for the kinetics model described in Figure 1. The model is not especially simple, nor is it completely well-defined, but that merely reflects the nature of the underlying physical phenomenon. Ice crystal growth involves the complex interplay of numerous molecular processes operating over a range of length and time scales, so developing a complete and quantitative physical understanding of the different growth regimes presents a significant challenge.

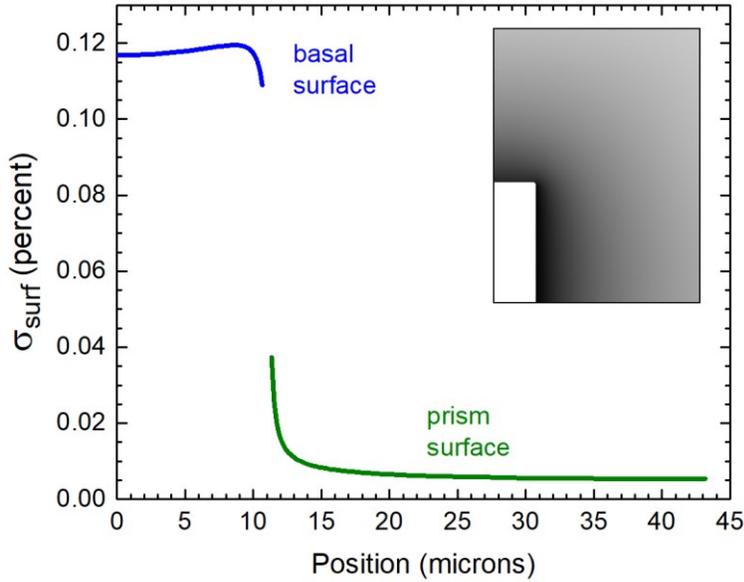

**Figure 13:** A model the supersaturation field surrounding the needle crystal described in Figures 11 and 12 at $t = 200$ seconds. The inset image illustrates the full numerical analysis in (r,z) space, showing a white ice crystal surrounded by a grey supersaturation field (with black equal to zero supersaturation). The lines show $\sigma_{surf}$ along the basal and prism surfaces in the cylindrically symmetrical model. The gap in the curves represents the basal/prism corner, which is slightly rounded by the Gibbs-Thomson effect. Although the details of this model are likely not completely accurate, it clearly indicates a large jump in $\sigma_{surf}$ between the basal and vicinal surfaces of the needle.

Comparing [2019Lib2] with the present paper, we see that the SDAK phenomenon that was so important on basal surfaces at -5 C is essentially absent at -2 C. The double-branched basal model at -5 C has been replaced by a remarkably simple basal curve at -2 C. Although the evidence is by no means conclusive at this point, the available experimental data supports an uncomplicated terrace-nucleation model for the growth of all basal surfaces at -2 C, as pictured in Figure 1.

In contrast, the growth of prism surfaces is relatively unchanged between -5 C and -2 C. In both cases, the slow growth of faceted surfaces is reasonably explained by a terrace nucleation model with $A_{prism} < 1$, while the growth of narrow prism facets on dendrite tips is better described by a separate branch on which $\alpha_{prism} \to 1$ at high $\sigma_{surf}$. I refer to the second branch in terms of "prism kinetic roughening," but the full nature of this behavior is not known at present.

Both the -2 C and -5 C models of the attachment kinetics are part of the full temperature-dependent model described in [2019Lib1]. In that paper I described a physical picture of the SDAK phenomenon as a function of temperature on both the basal and prism facets, while the following papers examined the experimental ramifications of the model in more detail at -5 C and -2 C. So far, the model seems to be holding up reasonably well to detailed quantitative comparisons with experimental data, so I am hopeful that the overall picture presented in [2019Lib1] is a significant step in the right direction.

One clear lesson from these investigations is that it is quite difficult to gain useful information pertaining to the attachment kinetics by analyzing ice growth data in normal air. Because $\alpha_{basal}$ and $\alpha_{prism}$ are both quite high in many experimental situations, the slow diffusion in air means $\alpha_{diff} \ll \alpha$, making it problematic to determine $\sigma_{surf}$ from growth measurements. The use of witness-surface analyses can provide some useful information, but the results are invariably somewhat model dependent. As described above, accurately reproducing the growth velocities of structures containing both vicinal and faceted surfaces, which includes most complex snow-crystal morphologies, will require rather sophisticated numerical modeling techniques.

Our overarching conclusion from these studies at -5 C and -2 C, however, is that we are able to reasonably explain a variety of disparate experimental results using a single model of the attachment kinetics. It is a somewhat complicated model overall, still not fully developed in spots, and the underlying physics is only partially understood. Nevertheless, despite its shortcomings, the model seems to provide a coherent, physically plausible picture of the ice/vapor attachment kinetics over a significant range of environmental conditions.